\documentclass[prx,twocolumn,english,superscriptaddress,floatfix,longbibliography]{revtex4-2}

\usepackage{graphicx}
\usepackage{dcolumn}
\usepackage{bm}

\usepackage[utf8]{inputenc}
\usepackage[T1]{fontenc}
\usepackage{mathptmx}
\usepackage{listings}
\usepackage{rotating} 
\usepackage{physics}

\usepackage{color}
\usepackage{dcolumn} 
\usepackage{bm} 
\usepackage{graphicx}
\usepackage{multirow} 
\usepackage{pifont} 
\usepackage{epsfig}
\usepackage{amsmath} 
\usepackage{subfigure}
\usepackage{float}
\usepackage{booktabs}
\usepackage{tabularx}
\usepackage{natbib}
\usepackage{textcomp}
\usepackage{gensymb}
\usepackage{rotating}
\usepackage[normalem]{ulem}

\begin{document}

\author{Joel D. Mallory}
\affiliation{
             Department of Chemistry and Biochemistry,
             Florida State University,
             Tallahassee, FL 32306-4390}

\author{A. Eugene DePrince III}
\email{adeprince@fsu.edu}
\affiliation{
             Department of Chemistry and Biochemistry,
             Florida State University,
             Tallahassee, FL 32306-4390}

\title{Reduced-density-matrix-based {\em ab initio} cavity quantum electrodynamics}

\begin{abstract}
A reduced-density-matrix (RDM)-based approach to {\em ab initio} cavity quantum electrodynamics (QED) is developed. The expectation value of the Pauli-Fierz Hamiltonian is expressed in terms of one- and two-body electronic and photonic RDMs, and the elements of these RDMs are optimized directly in polynomial time by semidefinite programming techniques, without knowledge of the full wave function. QED generalizations of important ensemble $N$-representability conditions are derived and enforced in this procedure. The resulting approach is applied to the description of classic ground-state strong electron correlation problems, augmented by the presence of ultrastrong light-matter coupling.  First, we assess cavity-induced changes to the singlet-triplet energy gap of the linear oligoacene series; for a heptacene molecule, this gap can change by as much as 1.9 kcal mol$^{-1}$ (or 15\%) when the molecule is aligned along the cavity mode polarization axis. We also explore the metal-insulator transition in linear hydrogen chains and demonstrate that strong electron-photon interactions increase the insulating character of these systems under all coupling strengths considered.
\end{abstract}

\maketitle

\section{Introduction}

Strong coupling of light and matter within optical micro/nano/picocavities\cite{Nori19_19,Narang18_1479,Barnes14_013901,Baumberg16_726,YuenZhou18_6325} can lead to the creation of hybrid light-matter states that exhibit interesting physicochemical properties that are often substantially different from those of the isolated states.\cite{Whittaker98_6697,Mugnier04_036404,Galitski11_490,Ebbesen12_1592,Ebbesen15_1123,Baumberg16_127,Ebbesen16_2403,Nitzan17_443003,Ebbesen17_9034,Bellessa19_173902,Forrest20_371,Owrutsky20_100902,YuenZhou18_6659,Feist19_8698,Sentef19_235156,Hsieh17_1077} The unique properties of these polariton states can be leveraged in a variety of chemical and physical contexts. For example, strong light-matter coupling can influence rates and selectivity of chemical\cite{Ebbesen16_11462,George19_10635,Ebbesen19_615} and photophysical\cite{Smith14_5561} processes, or it can be exploited to monitor\cite{Baumberg15_825} or suppress\cite{Shegai18_eaas9552} photochemical reactions.
Polariton formation is also often associated with exotic physical states and processes, including exciton-polariton condensates,\cite{Yamamoto08_700,Yamamoto14_803,Menon17_491} polariton lasing,\cite{Xiong17_3982,Rivas17_31} dynamical phase transitions,\cite{Thompson20_602} and even superfluidity at room temperature.\cite{Sanvitto17_837} Not surprisingly, such examples
have inspired the development of a variety of theoretical and computational tools with the hope that they can offer predictive insight into the fundamental principles that govern cavity-mediated processes.

Theoretical cavity QED studies have largely adopted simple model Hamiltonians\cite{Cummings63_89,Cummings68_379,Galbraith83_966,Eberly06_5621,Cao14_795,Pu16_053621} that can describe qualitative changes to optical properties or rates of reaction that result from strong interactions between cavity modes and few-level quantum emitters.\cite{Gray13_075411,Huo19_5519,Huo20_6321,YuenZhou18_167,Foley20_9063}
On the other hand, {\em ab initio} approaches potentially possess greater predictive power, particularly in the context of cavity-mediated changes to ground-state properties.\cite{Koch20_041043,DePrince21_094112,Flick22_4995,Koch21_094113} 
Consequently, a number of recent studies have used a first-principles description of the molecular system (via the Pauli-Fierz Hamiltonian\cite{Spohn04_book,Rubio18_0118}) in order to model interactions between electronic and photonic degrees of freedom in a rigorous and unbiased way.
Most of these efforts have been based on a quantum electrodynamical extension of density functional theory (QEDFT),\cite{Bauer11_042107,Rubio14_012508,Tokatly13_233001,Rubio17_3026,Rubio17_113036,Tokatly18_235123,Rubio15_093001,Rubio18_992,Rubio19_4883,Appel19_225,Narang20_094116,Rubio20_508,Narang21_104109,Shao21_064107,Shao22_124104,Rubio21_41} while others\cite{Koch20_041043,Manby20_023262,DePrince21_094112,DePrince22_2111,Flick22_4995,Koch21_094113,Flick21_9100,Koch22_234103} have taken a more sophisticated many-body approach rooted in coupled-cluster (CC)\cite{Cizek66_4256, Paldus71_359, Bartlett09_book, Musial07_291} theory. Additional {\em ab initio} cavity QED methods based on truncated\cite{Foley21_arxiv} or stochastic\cite{Varga21_273601}  configuration interaction (CI) have also been realized. While this field has experienced rapid and significant progress in recent years, it is notable that no {\em ab initio} cavity QED theories developed to date have specifically targeted the case of simultaneous strong electron--photon and strong electron--electron correlations. The purpose of this contribution is to address this gap and to develop a method that can describe both of these strong correlation effects in a computationally efficient way.

Strong correlations among electrons are notoriously difficult to describe, with the principal challenge being the steep computational cost of standard approaches for modeling these effects, such as the complete active space self-consistent field (CASSCF) method.\cite{Roos:1980:157,Siegbahn:1980:323,Siegbahn:1981:2384,Roos:1987:399} An exact treatment of the electronic structure of an active space increases exponentially with the size of this space, and, as a result, a method like CASSCF can only be applied to active spaces composed of at most 20 electrons distributed among 20 orbitals [a (20e,20o) active space].\cite{Vogiatzis17_184111} A QED-based extension of such an active space method would be similarly limited.
In this work, we circumvent this exponential complexity with a reduced density matrix (RDM) based approach to {\em ab initio} cavity QED, which, by design, does not require the storage or manipulation of the full polaritonic wave function. The method is a generalization of  variational two-electron RDM (2RDM) theory,\cite{Husimi:1940:264, Lowdin:1955:1474,Mayer:1955:1579,Rosina75_868,Rosina75_221,Garrod75_300,Rosina79_1366,Erdahl79_147,Fujisawa01_8282,Mazziotti02_062511,Mazziotti06_032501,Zhao07_553,Lewin06_064101,Bultinck09_032508,DePrince16_423,DeBaerdemacker11_1235,VanNeck15_4064,Mazziotti16_032516,DeBaerdemacker18_024105,Mazziotti17_084101,Ayers09_5558,Bultinck10_114113,Cooper11_054115,DePrince19_032509,Mazziotti16_153001,Mazziotti20_052819} which has been applied\cite{Mazziotti08_134108,Mazziotti11_5632,DePrince16_2260,DePrince19_6164,DePrince19_276} to active-space-based calculations of many-electron systems that are substantially larger than the practical limit of CASSCF mentioned above.
In the conventional (non-QED) variational 2RDM (v2RDM) approach, the 2RDM is determined by minimizing the electronic energy, which is a functional of the 2RDM, with respect to variations in its elements. Throughout this optimization, the 2RDM is constrained such that it satisfies approximate ensemble $N$-representability conditions,\cite{Coleman63_668,Percus64_1756,Erdahl78_697,Percus04_2095,Erdahl01_042113,Mazziotti12_263002} which are constraints meant to ensure that the 2RDM is derivable from an ensemble of $N$-electron density matrices. In the cavity QED generalization of v2RDM theory, additional photonic and mixed photon-electron RDMs appear in the energy expression, and the elements of these RDMs can also be obtained via a constrained optimization procedure. This procedure should enforce additional polaritonic representability conditions, and we derive an important subset of these conditions below.

The remainder of this work is organized as follows. The theoretical underpinnings of the RDM-based {\em ab initio} cavity QED approach, including important representability conditions that should be satisfied by polaritonic RDMs, are discussed in Sec.~\ref{SEC:THEORY}. Section \ref{SEC:COMPUTATION_DETAILS} then outlines the details of the QED-v2RDM calculations, the results of which are presented in Sec.~\ref{SEC:RESULTS}. QED-v2RDM is benchmarked for simple cases for which the theory should provide exact results, and it is then applied to elucidate cavity effects on ground states of two classic strong electron correlation problems. Specifically, we assess the effect that strong electron-photon coupling has on the singlet-triplet energy gap in the linear oligoacene series and the metal-insulator transition in linear hydrogen chains. Some concluding remarks can be found in Sec.~\ref{SEC:CONCLUSIONS}.

\section{Theory}

\label{SEC:THEORY}
In this section, we describe the Pauli-Fierz Hamiltonian for {\em ab initio} cavity QED and provide an overview of the QED-v2RDM approach for determining the ground-state energy associated with this operator. Throughout, the lower-case labels $p$, $q$, $r$, and $s$ refer to orthonormal electronic spin orbitals, capital labels $A$ and $B$ refer to photon-number states, and non-italicized superscripts ``e'' and ``p'' refer to electronic and photonic degrees of freedom, respectively.  

\subsection{The Pauli-Fierz Hamiltonian}
Consider a single molecule or molecular ensemble confined to an optical cavity that supports a single photon mode. A first-principles description of the interactions between the electronic and the photonic degrees of freedom can be obtained from the Pauli-Fierz Hamiltonian, $\hat{H}_{\rm PF}$,\cite{Spohn04_book,Rubio18_1} which we express in the length gauge and under the dipole and cavity Born-Oppenheimer approximations:
\begin{eqnarray}
    \label{eq:Hfull}
    \hat{H}_{\rm PF}&=&\hat{H}_{\rm e}+\omega_{\rm cav}\hat{b}^{\dagger}\hat{b} -\sqrt{\frac{\omega_{\rm cav}}{2}}(\bm{\lambda} \vdot \bm{\mu})(\hat{b}^{\dagger}+\hat{b}) \nonumber \\ 
    &+&\frac{1}{2}(\bm{\lambda} \vdot \bm{\mu})^2
\end{eqnarray}   
The first term in Eq.~\ref{eq:Hfull} is purely electronic and has the familiar form
\begin{equation}
    \label{eq:Helect1}
    \hat{H}_{\rm e}=\sum_{pq}(T_{pq} + V_{pq})\hat{a}_p^{\dagger}\hat{a}_q+\frac{1}{2}\sum_{pqrs}\langle pq|rs\rangle \hat{a}_p^{\dagger}\hat{a}_q^{\dagger}\hat{a}_s\hat{a}_r
\end{equation}
where $T_{pq}$ and $V_{pq}$ are electron kinetic energy and electron-nuclear attraction integrals, respectively, and $\langle pq|rs\rangle$ denotes a two-electron repulsion integral in physicists' notation. The symbols $\hat{a}_p^{\dagger}$ and $\hat{a}_p$ represent fermionic creation and annihilation operators, respectively, which act within the electronic space. The second term in Eq.~\ref{eq:Hfull} is purely photonic and describes the energy levels of a cavity mode with fundamental frequency, $\omega_{\rm cav}$. The symbols $\hat{b}^{\dagger}$ and $\hat{b}$ represent bosonic creation and annihilation operators that create and destroy one photon in the cavity mode, respectively. The third term describes dipolar coupling between the molecule and the cavity. Here, $\bm{\mu}$ represents the total molecular dipole operator (nuclear plus electronic), and $\bm{\lambda}$ is a coupling vector that parametrizes the strength of the photon-electron interactions (and is related to the cavity mode volume; see Eq.~\ref{eq:lambda}). Finally, the last term in Eq.~\ref{eq:Hfull} represents the  dipole self-energy, which arises when the Pauli-Fierz Hamiltonian is represented within the length gauge.\cite{Rubio18_034005} As a matter of convenience, we also choose to transform $\hat{H}_{\rm PF}$ to the coherent-state basis,\cite{Koch20_041043} after which it takes the form
\begin{eqnarray}
    \label{eq:Hfullcoherent}
    \hat{H}_{\rm PF}&=&\hat{H}_{\rm e}+\omega_{\rm cav}\hat{b}^{\dagger}\hat{b}-\sqrt{\frac{\omega_{\rm cav}}{2}}(\bm{\lambda} \vdot [\bm{\mu} -\langle \bm{\mu} \rangle])(\hat{b}^{\dagger}+\hat{b})\nonumber \\
    &+&\frac{1}{2}(\bm{\lambda} \vdot [\bm{\mu}-\langle \bm{\mu} \rangle])^2
\end{eqnarray}
Here, the symbol $\langle \bm{\mu} \rangle$ represents the expectation value of the total dipole operator with respect to a reference state, which we take to be the direct product of a Slater determinant of electronic orbitals, $|0^{\rm e}\rangle$ and a zero-photon photon-number state, $|0^{\rm p}\rangle $ ({\em i.e.}, the QED Hartree-Fock [QED-HF] state).

\subsection{Electronic and photonic reduced density matrices}
The most general correlated polaritonic wave function describing the state of a many-electron system coupled to a single-mode cavity is the full configuration interaction wave function of the form
\begin{equation}
\label{EQN:FCI}
    |\Psi\rangle = \sum_{\mu A} c_{\mu A} | \mu^{\rm e}\rangle | A^{\rm p}\rangle
\end{equation}
Here, $|\mu^{\rm e}\rangle$ represents a Slater determinant of electronic orbitals, $|A^{\rm p}\rangle$ is a photon-number state representing $A$ photons in the cavity mode, and $c_{\mu A}$ is an expansion coefficient. This wave function contains far more information than is necessary to define the ground-state energy of the Pauli-Fierz Hamiltonian. This is because $\hat{H}_{\rm PF}$ involves only one- and two-body interactions, and its expectation value is exactly expressible with knowledge of only one- and two-body reduced density matrices (RDMs). For example, consider the one-electron RDM (${}^1${\bf D}$_{\rm e}$) and the two-electron RDM (${}^2${\bf D}$_{\rm ee}$), the elements of which are defined as 
\begin{equation}
\label{EQN:eRDM}
    ^1D^p_q=\langle\Psi|\hat{a}_p^{\dagger}\hat{a}_q|\Psi\rangle
\end{equation}
and
\begin{equation}
\label{EQN:eeRDM}
    ^2D^{pq}_{rs}=\langle\Psi|\hat{a}_p^{\dagger}\hat{a}_q^{\dagger}\hat{a}_s\hat{a}_r|\Psi\rangle
\end{equation} 
respectively. With these definitions, it can be seen that the expectation value of the electronic Hamiltonian, $\hat{H}_{\rm e}$, reduces to a functional of these quantities:
\begin{equation}
  \label{eq:Eelect}
  \langle \Psi | \hat{H}_{\rm e} | \Psi \rangle = \sum_{pq}(T_{pq}+V_{pq}) {^1D}^p_q+\frac{1}{2}\sum_{pqrs}\langle pq|rs\rangle {^2D}^{pq}_{rs}
\end{equation}
Similarly, because the dipole self-energy terms include up to two-electron interactions, the expectation value of that quantity is also expressible in terms of the elements of these RDMs. 

The expectation values of the remaining terms in Eq.~\ref{eq:Hfullcoherent} are expressible in terms of photon and photon-electron RDMs. We arrive at the relevant expressions by first defining a photon density operator that connects photon-number states
\begin{equation}
    \hat{D}_{AB} = |A^{\rm p} \rangle \langle B^{\rm p}|
\end{equation}
and a photon RDM (${}^1${\bf D}$_{\rm p}$) that has elements equal to the expectation value of this operator
\begin{equation}
\label{EQN:pRDM}
    {}^1D^A_B = \langle \Psi | \hat{D}_{AB} |\Psi \rangle
\end{equation}
If we express the boson creation and annihilation operators as
\begin{equation}
    \hat{b}^\dagger = \sum_{A} \sqrt{A+1} |(A+1)^{\rm p}\rangle \langle A^{\rm p}|,
\end{equation}
and
\begin{equation}
    \hat{b} = \sum_{A} \sqrt{A} |(A-1)^{\rm p}\rangle \langle A^{\rm p}|
\end{equation}
respectively, it becomes clear that the expectation value of the bare cavity Hamiltonian (the second term in Eq.~\ref{eq:Hfullcoherent}) is expressible in terms of the diagonal elements of the photon RDM as
\begin{equation}
    \omega_{\rm cav}\langle \Psi | \hat{b}^\dagger \hat{b} | \Psi \rangle = \omega_{\rm cav} \sum_A A {}^1D^A_A
\end{equation}

The expectation value of the bilinear coupling term (the third term in Eq.~\ref{eq:Hfullcoherent}) involves the off-diagonal elements of the photon RDM as well as an additional electron-photon RDM (${}^2${\bf D}$_{\rm ep}$), the elements of which are defined as
\begin{equation}
\label{EQN:epRDM}
    {}^2D^{pA}_{qB} = \langle \Psi| \hat{a}^\dagger_p  \hat{a}_q \hat{D}_{AB}|\Psi \rangle
\end{equation}
Specifically, we have
\begin{eqnarray}
     \langle \Psi | (\bm{\lambda} \vdot [\bm{\mu}-&\langle \bm{\mu} \rangle&])(\hat{b}^{\dagger}+\hat{b}) | \Psi \rangle = \nonumber \\ 
     \sum_{\xi} \lambda_\xi \bigg ( \sum_{pq} \mu_{pq}^\xi &\sum_A& \bigg [\sqrt{A+1}~{}^2D^{p(A+1)}_{qA} + \sqrt{A} ~{}^2D^{p(A-1)}_{qA} \bigg ]  \nonumber \\
     -  \langle {\bm \mu}_{\rm e} \rangle_\xi &\sum_A& \bigg [ \sqrt{A+1}~{}^1D^{A+1}_A + \sqrt{A}~{}^1D^{A-1}_A \bigg ]  \bigg )
\end{eqnarray}
where $\lambda_\xi$ represents a component of the coupling vector (with $\xi \in \{x,y,z\}$), $\langle {\bm \mu}_{\rm e} \rangle_\xi$ represents the $\xi$-component of the expectation value of the electronic dipole operator (with respect to the QED-HF reference configuration), and $\mu_{pq}^{\xi}$ represents an integral over the $\xi$-component of the electronic dipole operator. Note that the nuclear contributions to the molecular dipole vanish after taking the expectation value with respect to $|\Psi\rangle$. 

Now, because we have shown that the expectation value of the Pauli-Fierz Hamiltonian is expressible in terms of one- and two-body RDMs, our goal is to determine the elements of these RDMs directly by minimizing the total energy with respect to their elements. However, we must impose a variety of conditions on these RDMs in order to guarantee that they correspond to a physically meaningful polaritonic wave function. Necessary conditions enforced in this optimization procedure are outlined below.

\subsection{Polaritonic representability conditions}
An electronic RDM that is derivable from an ensemble of $N$-electron density matrices is said to be ensemble $N$-representable,\cite{Coleman63_668} and
a large body of work\cite{Rosina75_868,Rosina75_221,Garrod75_300,Rosina79_1366,Erdahl79_147,Fujisawa01_8282,Mazziotti02_062511,Mazziotti06_032501,Zhao07_553,Lewin06_064101,Bultinck09_032508,DePrince16_423,DeBaerdemacker11_1235,VanNeck15_4064,DeBaerdemacker18_024105,Mazziotti17_084101,Ayers09_5558,Bultinck10_114113,Cooper11_054115,DePrince19_032509,Mazziotti08_134108,DePrince16_2260,Mazziotti16_153001} describes the direct variational determination of the two-electron RDM, subject to known, necessary $N$-representability constraints.\cite{Percus64_1756,Erdahl78_697,Percus04_2095,Erdahl01_042113,Mazziotti12_263002} In this section, we generalize the concept of $N$-representability to the cavity QED case and develop polaritonic representability conditions that resemble the two-particle (PQG) constraints of Garrod and Percus.\cite{Percus64_1756} It should be noted that, as in standard electronic structure theory, the polaritonic generalizations of the PQG constraints are, in general, necessary yet insufficient for guaranteeing the ensemble $N$-representability of the optimized RDMs.

We begin by discussing basic statistical conditions that should be satisfied by electronic and photonic RDMs. First, consider that $|\Psi\rangle$ should be an eigenfunction of the electronic number operator, $\hat{N}$, 
\begin{equation}
\label{EQN:NUMBER}
    \hat{N}|\Psi\rangle = \sum_{p} \hat{a}^\dagger_p \hat{a}_p |\Psi \rangle = N| \Psi \rangle
\end{equation}
This relationship implies multiple constraints on the two-electron RDM, including a constraint on its trace
\begin{equation}
    \label{eq:electpairs}
    \sum_{pq} {}^2D_{pq}^{pq}=N(N-1)
\end{equation}
and contraction constraints that relate the two-electron RDM to the one-electron RDM
\begin{equation}
    \label{eq:12relate}
    (N-1){}^1D_q^p=\sum_r {}^2D_{qr}^{pr},
\end{equation}
and the electron-photon RDM to the photon RDM
\begin{equation}
    \label{eq:12relatephoton}
    N{^1D}_B^A=\sum_r {^2D}_{rB}^{rA}
\end{equation}
We can devise additional constraints from spin considerations. First, an eigenfunction of the square of the spin angular momentum operator, $\hat{S}^2$, should satisfy
\begin{equation}
\label{EQN:S2}
     \hat{S}^2 | \Psi \rangle = S(S+1)| \Psi \rangle
\end{equation}
which implies an additional constraint on the two-electron RDM.\cite{Mazziotti05_052505}
Second, the Pauli-Fierz Hamiltonian is non-relativistic, so $|\Psi\rangle$ could also be an eigenfunction of the $z$-component of the spin operator, $\hat{S}_z$. Enforcing $\hat{S}_z$ symmetry leads to trace and contraction constraints that are similar to those above, except that they apply to the individual spin-blocks of the two-electron and electron-photon RDMs.
Lastly, the photon-number states form a complete set within the photonic space, which implies the identity $\sum_A |A\rangle \langle A| = 1$, and a constraint on the trace of the photon RDM follows:
\begin{equation}
   \sum_A {}^1D^A_A = 1
\end{equation}

More complex polaritonic representability constraints can be derived following the strategy outlined in Ref.~\citenum{Erdahl01_042113}. Consider $|\Psi\rangle$, as parametrized in Eq.~\ref{EQN:FCI}. A set of operators, $\hat{C}^\dagger_I$, can be used to generate a basis of functions out of this state as
\begin{equation}
    |\phi_I\rangle = \hat{C}^\dagger_I | \Psi \rangle
\end{equation}
and the Gram matrix associated with this basis must be non-negative. Put another way, for a matrix $M$ with elements $M_{IJ} = \langle \phi_I | \phi_J \rangle$, we have
\begin{equation}
    M \succeq 0
\end{equation}
Different choices for $\hat{C}_I^\dagger$ imply the non-negativity of the eigenvalues of various RDMs. For example, for $\hat{C}^\dagger_I = \hat{a}_q$, we find that the one-electron RDM (${}^1${\bf D}$_{\rm e} \succeq 0$) should be non-negative, and the choice $\hat{C}^\dagger_I = \langle B^{\rm p}|$ implies the same property for the photon RDM (${}^1${\bf D}$_{\rm p} \succeq 0$). Table \ref{TAB:POSITIVITY} details the choices for $\hat{C}^\dagger_I$ that lead to the polaritonic generalization of the PQG constraints that are familiar in electronic structure theory. 
\begin{table}[!htpb]
    \caption{Generators for polaritonic representability conditions involving one-, two-, and three-body RDMs.}
    \label{TAB:POSITIVITY}
    \begin{center}
    \begin{tabular}{l l l}
    \hline\hline
    $\hat{C}^\dagger_I$ & RDM type & RDM definition \\
    \hline
    $\hat{a}_q$           & one-electron & ${}^1D^p_q = \langle \Psi | \hat{a}^\dagger_p \hat{a}_q | \Psi \rangle$ \\
    $\hat{a}^\dagger_q$   & one-hole     & ${}^1Q^p_q = \langle \Psi | \hat{a}_p
    \hat{a}^\dagger_q | \Psi \rangle$ \\
    $\hat{a}_s\hat{a}_r$           & two-electron & ${}^2D^{pq}_{rs} = \langle \Psi | \hat{a}^\dagger_p \hat{a}^\dagger_q\hat{a}_s\hat{a}_r | \Psi \rangle$ \\
    $\hat{a}^\dagger_s\hat{a}^\dagger_r$           & two-hole & ${}^2Q^{pq}_{rs} = \langle \Psi | \hat{a}_p \hat{a}_q\hat{a}^\dagger_s\hat{a}^\dagger_r | \Psi \rangle$ \\
    $\hat{a}^\dagger_s\hat{a}_r$           & electron-hole & ${}^2G^{pq}_{rs} = \langle \Psi | \hat{a}^\dagger_p \hat{a}_q\hat{a}^\dagger_s\hat{a}_r | \Psi \rangle$ \\
    $\langle B^{\rm p}|$  & photon &     ${}^1D^A_B = \langle \Psi | \hat{D}_{AB} | \Psi \rangle$ \\
    $\langle B^{\rm p}|\hat{a}_q$ & electron-photon & ${^2D}_{pB}^{qA}=\langle\Psi|\hat{a}_p^{\dagger}\hat{D}_{AB} \hat{a}_q|\Psi\rangle$ \\
    $\langle B^{\rm p}|\hat{a}_s\hat{a}_r$ & electron-electron-photon & ${^3D}_{rsB}^{pqA}=\langle\Psi|\hat{a}_p^{\dagger}\hat{a}_q^{\dagger}\hat{D}_{AB} \hat{a}_s\hat{a}_r|\Psi\rangle$ \\
    $\langle B^{\rm p}|\hat{a}^\dagger_s\hat{a}^\dagger_r$ & hole-hole-photon & ${^3Q}_{rsB}^{pqA}=\langle\Psi|\hat{a}_p\hat{a}_q\hat{D}_{AB} \hat{a}^{\dagger}_s\hat{a}^{\dagger}_r|\Psi\rangle$ \\
    $\langle B^{\rm p}|\hat{a}^\dagger_s\hat{a}_r$ & electron-hole-photon & ${^3G}_{rsB}^{pqA}=\langle\Psi|\hat{a}_p^{\dagger}\hat{a}_q\hat{D}_{AB} \hat{a}^{\dagger}_s\hat{a}_r|\Psi\rangle$ \\
    \hline
 
\hline\hline
\end{tabular}
\end{center}
\end{table}
Note that, in their conventional form (which was derived for the electronic problem), the PQG constraints involve at most two-body quantities, but here, we consider $\hat{C}^\dagger_I$ involving up to two fermionic operators and one bosonic operator, which leads to constraints on three-body RDMs. These three-body RDMs can be related to lower-order RDMs through symmetries that should be preserved by the wave function from which they derive. For example,  Eq.~\ref{EQN:NUMBER} implies that the electron-electron-photon RDM defined in Table \ref{TAB:POSITIVITY} should contract to the two-body electron-photon RDM according to 
\begin{equation}
    \label{eq:123interrelate}
    (N-1) {^2D}_{qB}^{pA}=\sum_r {^3D}_{qrB}^{prA}
\end{equation}
Enforcing $\hat{S}_z$ symmetry again leads to additional contraction constraints connecting the spin-blocks of the electron-electron-photon and electron-photon RDMs, and Eq.~\ref{EQN:S2} implies additional constraints that relate the electron-electron-photon RDM and the photon RDM. Additional details regarding spin-related constraints are provided in Appendix A.

\section{Computational Details}
  
\label{SEC:COMPUTATION_DETAILS}

We determine the elements of the RDMs defined in Table \ref{TAB:POSITIVITY} by minimizing the expectation value of the Pauli-Fierz Hamiltonian with respect to the elements of the RDMs, while also constraining the RDMs to be positive semidefinite and to satisfy the linear mappings and expectation value constraints outlined in the previous section. This constrained energy optimization can be cast as a semidefinite program (SDP), the primal form of which is
\begin{eqnarray}
\operatorname{minimize} \quad &E_{\rm primal}=\mathbf{c}^{T} \mathbf{x}\\
\operatorname { such\ that } \quad &\mathbf{A x}=\mathbf{b}\\
\operatorname { and } \quad &M(\mathbf{x}) \succeq 0
\end{eqnarray}
Here {\bf x} represents the primal solution vector, and {\bf c} represents a vector containing the one- and two-electron integrals that define the system. The elements of {\bf x} map onto the RDMs defined in Table \ref{TAB:POSITIVITY} via the mapping $M(\mathbf{x})$, and the notation $M(\mathbf{x}) \succeq 0$ indicates that the RDMs are all positive semidefinite. The elements of {\bf c} are arranged such that the primal energy, $E_{\rm primal}$, is equal to the expectation value of the Pauli-Fierz Hamiltonian. Lastly, the symbols {\bf A} and {\bf b} represent the constraint matrix and vector, respectively, which encode the linear mappings between RDMs and the expectation value constraints.
We implemented the QED-v2RDM model as an SDP in a locally modified version of \texttt{hilbert},\cite{hilbert1} which is a plugin to the \textsc{Psi4} electronic structure package.\cite{Sherrill20_184108}  This plugin contains an interface to a library of SDP solvers, \texttt{libsdp}, which includes algorithms for the solving the SDP problem that are similar to those described in Refs.~\citenum{Mazziotti04_213001} and \citenum{Mazziotti11_083001}. 

QED-v2RDM calculations considered two photon-number states, corresponding to zero or one photon in the cavity. As such,  for two-electron and two-hole systems, QED-v2RDM performed under the cavity QED generalization of the PQG conditions should be numerically equivalent to a QED generalization of CC (QED-CC) that considers single and double electron excitations plus single photon transitions (QED-CCSD-1, see Ref.~\citenum{Koch20_041043}). To verify the correctness of our algorithm, we performed benchmark calculations on cavity-embedded molecular hydrogen and hydrogen fluoride using the cc-pVDZ and STO-3G basis sets, respectively, at the QED-v2RDM and QED-CCSD-1 levels of theory. The QED-CCSD-1 calculations were carried out using the implementation available in \texttt{hilbert}. All calculations used the density-fitting (DF) approximation to the electron repulsion integrals (ERIs),\cite{Whitten73_4496,Sabin79_3396} with the cc-pVDZ-JKFIT auxiliary basis set. QED-v2RDM calculations on H$_2$ and hydrogen fluoride were considered converged when the primal-dual energy gap fell below $1\times 10^{-6}$ E$_{\rm h}$ and the primal and dual representability errors fell below $1\times 10^{-6}$. For additional details on how these quantities are defined, the reader is referred to Ref.~\citenum{DePrince19_6164}.

The QED-v2RDM approach was applied to two classic strong correlation problems. First, we performed active-space QED-v2RDM calculations on the linear oligoacene series (or $k$-acene, C$_{4k+2}$H$_{2k+4}$, with $k$=3--7), with all molecules described by the cc-pVDZ basis set. All calculations used the DF approximation to the ERIs, with the cc-pVDZ-JKFIT auxiliary basis set. Geometries for the oligoacene molecules, which were taken from Ref.~\citenum{DePrince19_6164}, were optimized at the v2RDM-CASSCF level of theory. The oligoacene active spaces were composed of the valence $\pi$-network with $4k+2$ electrons distributed among $4k+2$ orbitals [\textit{i.e.}, ([$4k+2$]e, [$4k+2$]o) active spaces; see Ref.~\citenum{DePrince19_6164} for additional details]. Active-space QED-v2RDM calculations were performed using the same ([$4k+2$]e, [$4k+2$]o) active spaces, but the present computations do not take advantage of spatial symmetry, while those in Ref.~\citenum{DePrince19_6164} did. We also note that we do not optimize the orbitals as part of the active-space QED-v2RDM procedure; the orbitals are taken from QED-HF calculations.  With the exception of 7-acene, active-space QED-v2RDM calculations for the oligoacene molecules were considered converged when the primal-dual energy gap fell below $1\times 10^{-5}$ E$_{\rm h}$ and the primal and dual representability errors fell  below $1\times 10^{-4}$. The reported singlet and triplet energies for 7-acene were obtained from calculations in which the primal-dual energy gap and the representability errors were converged to $5 \times 10^{-4}$ E$_{\rm h}$ and $5 \times 10^{-5}$, respectively. The reported natural orbital occupation numbers for 7-acene were obtained from calculations where these quantities were converged to $1\times 10^{-2}$ E$_{\rm h}$ and $5\times 10^{-4}$, respectively. Second, we performed QED-v2RDM calculations on linear hydrogen chains within the STO-3G basis set. All calculations on hydrogen chains used Cholesky-decomposed ERIs with a tight decomposition threshold of $1\times 10^{-12}$ E$_{\rm h}$, and the QED-v2RDM calculations were considered converged when the primal-dual energy gap and primal and dual representability errors fell below $1\times 10^{-5}$ E$_{\rm h}$ and $1\times 10^{-5}$, respectively.

Lastly, in all QED-v2RDM and QED-CCSD-1 calculations performed in this work, we consider cavity modes polarized along a single Cartesian direction [{\em e.g.}, $\bm \lambda = (0, 0, \lambda)$]. In the limit of single-molecule coupling, $\lambda$ can be related to an effective cavity mode volume by\cite{Feist19_021057}
\begin{equation}
    \label{eq:lambda}
    \lambda=\sqrt{\frac{1}{\epsilon_0 V_{\rm eff}}}
\end{equation}
where $\epsilon_0$ represents the permittivity of free-space. The largest coupling strength that we consider is $\lambda=0.05$ atomic units, which corresponds to an effective mode volume of approximately 0.74 nm$^3$. This small effective volume can be achieved experimentally in so-called picocavity setups.\cite{Baumberg16_726,Baumberg18_7146} We note, however, that such a small effective mode volume may result in significant inhomogeneity of optical fields\cite{Aizpurua18_2358} and that the present model does not account for such effects. One could, in principle, replace the single cavity mode for such cases with a collection of quantized field modes characterized by different coupling strengths that account for spatial information ({\em i.e.}, the polarization of the mode plus its position relative to the molecular orbitals). Model Hamiltonians incorporating such spatial information have been considered elsewhere in the context of mean-field\cite{DePrince15_214104} and coupled-cluster\cite{Corni21_6664} approaches, and the present QED-v2RDM approach could also be generalized in this way. Assuming the quantized field modes are non-interacting, the cost of such a QED-v2RDM calculation would increase roughly linearly with the number of modes owing to the need to consider multiple fermion-photon RDMs.

\section{Results and Discussion}

\label{SEC:RESULTS}

In this Section, we benchmark QED-v2RDM against QED-CCSD-1 for two cases for which both theories should provide equivalent results: two-electron and two-hole systems coupled to a single-mode optical cavity. QED-v2RDM is then applied to the singlet-triplet energy gap in the oligoacene series, as well as to the metal-insulator transition in linear hydrogen chains. 

\subsection{Benchmarking QED-v2RDM}

\begin{figure*}[!htb]
    \centering
    \includegraphics{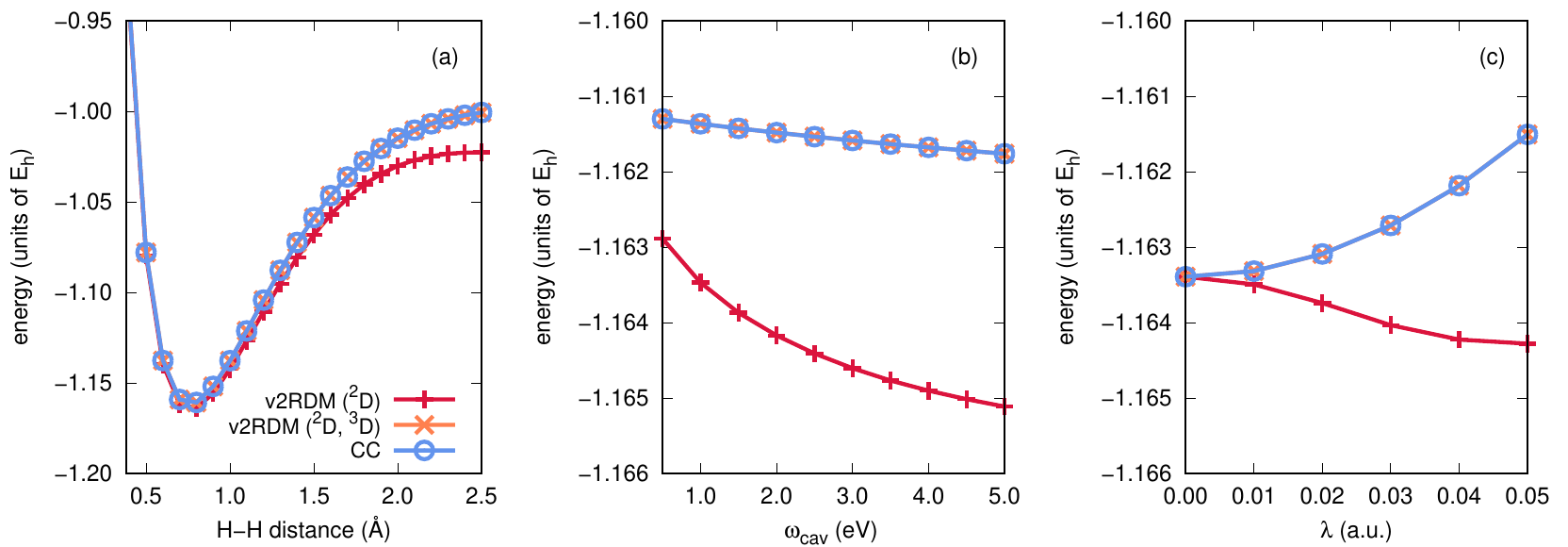}
    \caption{Potential energy curves (units of E$_{\rm h}$) as a function of (a) interatomic distance, (b) cavity mode frequency, and (c) cavity coupling strength for the lowest-energy singlet state of molecular hydrogen described by the cc-pVDZ basis set.}
    \label{FIG:H2SINGLET}
\end{figure*}

For two-electron and two-hole systems, QED-CCSD-1 is equivalent to a full CI performed using the expansion given in Eq.~\ref{EQN:FCI}, if that expansion is limited to consider at most single-photon transitions. In other words, QED-CCSD-1 provides an exact description of the polaritonic structure for such systems, under the assumption that higher-than-single photon transitions can be ignored. Under the two-particle (PQG) $N$-representability conditions detailed in Table \ref{TAB:POSITIVITY}, QED-v2RDM should also recover full CI results for these cases, so comparisons to QED-CCSD-1 provide a useful measure of the correctness of our implementation. Such comparisons also highlight the necessity of three-body $N$-representability conditions for an exact QED-v2RDM-based description of cavity-bound two-electron and two-hole systems.

Figure \ref{FIG:H2SINGLET} illustrates potential energy curves for the lowest-energy singlet state of a two-electron system, H$_2$, evaluated at the QED-CCSD-1 and QED-v2RDM levels of theory, using the cc-pVDZ basis set. The molecule is coupled to a single-mode cavity with the cavity mode polarization axis parallel to the molecular axis. Panel (a) of Fig.~\ref{FIG:H2SINGLET} depicts the energy as a function of the H--H distance, with cavity parameters $\omega_{\rm cav}$ and $\lambda$ fixed at 2.214 eV (or a wavelength of 560 nm) and 0.05 atomic units, respectively. Panel (b) depicts a scan over $\omega_{\rm cav}$, with fixed H--H distance (0.74 \AA) and fixed coupling strength ($\lambda = 0.05$ atomic units). Lastly, panel (c) provides energies computed at various coupling strengths, with fixed H--H distance (0.74 \AA) and fixed cavity mode frequency (2.214 eV). 
Two curves in each panel are labeled v2RDM, and these curves differ in the $N$-representability conditions applied within the computations. The label ``v2RDM (${}^2$D)'' indicates that the calculations were carried out under conditions that depend upon up to two-body RDMs (specifically, the one-electron, one-hole, two-electron, photon,  and electron-photon RDMs), while the label ``v2RDM (${}^2$D, ${}^3$D)'' indicates the application of the same conditions plus those that depend on the electron-electron-photon RDM. The definitions of these RDMs can be found in Table \ref{TAB:POSITIVITY}. 

Given that cavity-bound H$_2$ is a three-body system (two electrons plus a photon degree of freedom), we expect QED-v2RDM calculations that consider the non-negativity of the electron-electron-photon RDM to be the exact ones. Indeed, curves derived from QED-v2RDM calculations carried out with these three-body conditions are numerically indistinguishable from those obtained at the QED-CCSD-1 level of theory. On the other hand, calculations that consider only two-body constraints result in significantly lower energies; for example, as can be seen in Fig.~\ref{FIG:H2SINGLET}(a), v2RDM (${}^2$D) yields an energy 0.022 E$_{\rm h}$ lower than the exact energy at an H--H distance of 2.5 \AA.   Moreover, the v2RDM (${}^2$D) energies in Fig.~\ref{FIG:H2SINGLET}(c) display qualitatively incorrect behavior with increasing coupling strength. These observations are consistent with the behavior of conventional (non-QED) v2RDM theory; the v2RDM energy is a lower-bound to the full CI one, and over-correlation such as that exhibited by the v2RDM (${}^2$D) curves here indicates that the RDMs violate important $N$-representability conditions.  In this case, v2RDM (${}^2$D) performs poorly because the resulting two-electron and electron-photon RDMs are not derivable from a non-negative electron-electron-photon RDM. A similar analysis has confirmed the numerical equivalence of QED-v2RDM and QED-CCSD-1 for the lowest-energy triplet state of H$_2$, provided that QED-v2RDM calculations consider the non-negativity of the electron-electron-photon RDM; the results of this study can be found in Appendix B.

Figure \ref{FIG:HFSINGLET} illustrates potential energy curves for the lowest-energy singlet state of a two-hole system, hydrogen fluoride represented by the STO-3G basis set, evaluated at the QED-CCSD-1 and QED-v2RDM levels of theory.  As above, this molecule is coupled to a single-mode cavity with the cavity mode polarized along the molecular axis. Panels (a)-(c) of Fig.~\ref{FIG:HFSINGLET} are structured similarly to the same panels in Fig.~\ref{FIG:H2SINGLET}, except that the H--F distance in panels (b) and (c) is fixed at 0.917 \AA. In analogy to Fig.~\ref{FIG:H2SINGLET}, two curves in each panel are labeled QED-v2RDM, and these curves differ in the $N$-representability conditions applied within the computations. 
For a two-hole-plus-one-photon system such as this one, $N$-representability conditions involving the two-hole RDM (${}^2$Q) and the hole-hole-photon RDM (${}^3$Q) are expected to play energetically important roles (see Table \ref{TAB:POSITIVITY} for the definitions of these RDMs).
The label ``v2RDM (${}^2$Q)'' indicates that the calculations were carried out under conditions that depend upon all two-body RDMs given in Table \ref{TAB:POSITIVITY}, except for the particle-hole RDM (${}^2$G), plus conditions that depend upon the electron-electron-photon RDM. The label ``v2RDM (${}^2$Q, ${}^3$Q)'' indicates the application of the same conditions plus those that depend on the hole-hole-photon RDM. Essentially the same conclusions can be drawn here as were drawn above. QED-v2RDM can be numerically equivalent to QED-CCSD-1 for two-hole cavity-bound systems, provided that one considers $N$-representability conditions associated with the hole-hole-photon RDM. When such conditions are ignored, QED-v2RDM-derived energies are significantly lower than exact ones; the largest deviation between the exact energy and that from QED-v2RDM (${}^2$Q) is -0.020 E$_{\rm h}$ at an H--F distance of 2.5 \AA~[Fig.~\ref{FIG:HFSINGLET}(a)]. 

In light of these results for cavity-bound two-electron and two-hole systems, it is apparent that a QED generalization of the two-body $N$-representability conditions for electronic structure theory should consider three-body RDMs associated with electron-electron, hole-hole, and electron-hole pairs plus a photon degree of freedom. For the remainder of this work, $N$-representability conditions associated with all of the one-, two-, and three-body RDMs tabulated in Table \ref{TAB:POSITIVITY} are applied in all calculations.

\begin{figure*}[!htpb]
    \centering
    \includegraphics{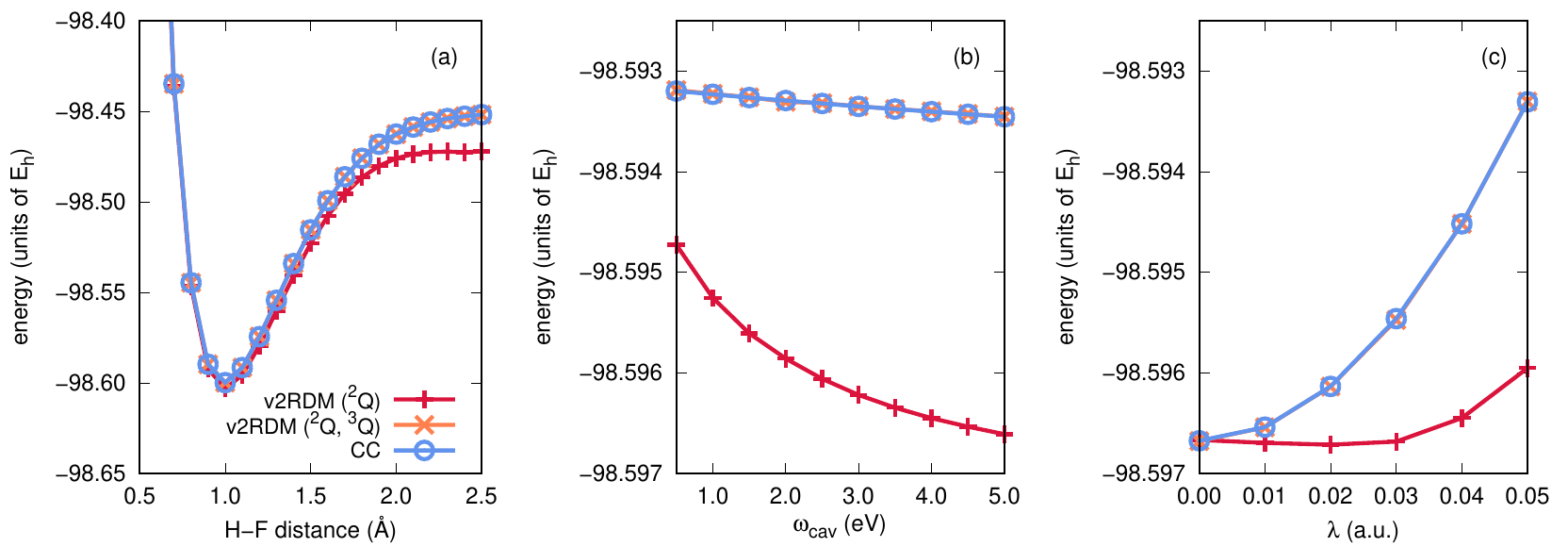}
    \caption{Potential energy curves (units of E$_{\rm h}$) as a function of (a) interatomic distance, (b) cavity mode frequency, and (c) cavity coupling strength for hydrogen fluoride described by the STO-3G basis set.}
    \label{FIG:HFSINGLET}
\end{figure*}

\subsection{Singlet-triplet energy gaps of oligoacene molecules}

We now use QED-v2RDM to explore cavity-induced changes to the electronic structure of the linear oligoacene (or $k$-acene) series. These molecules are known to have desirable optoelectronic properties, including the ability of some members of the series to undergo singlet fission,\cite{Schmidt15_2367,Nozik06_074510,HeadGordon11_19944,Stienkemeier17_2068,Friend13_1330} and theoretical studies suggest that singlet fission rates or yields can be influenced by coupling the molecules to optical cavity modes.\cite{YuenZhou18_1951,Mukamel21_2052} However, obtaining a reliable first-principles description of the electronic structure of oligoacene molecules can be challenging, especially for the longer members of the series whose ground-states are characterized by strong electron correlations and complex singlet polyradical character.\cite{Chan07_134309} As a result, these molecules have become a standard test for benchmarking the performance of methods designed for the strong correlation problem.\cite{DePrince16_2260,Mazziotti08_134108,Yanai13_401,Leininger17_3746,HeadGordon17_602,HeadGordon18_547,Casula18_134112,Evangelista16_161106,DePrince19_290} Here, we apply QED-v2RDM to the lowest-energy singlet and triplet states of cavity-coupled $k$-acene molecules composed of up to seven fused six-membered rings. 

\begin{figure}[!htb]
    \centering
    \includegraphics{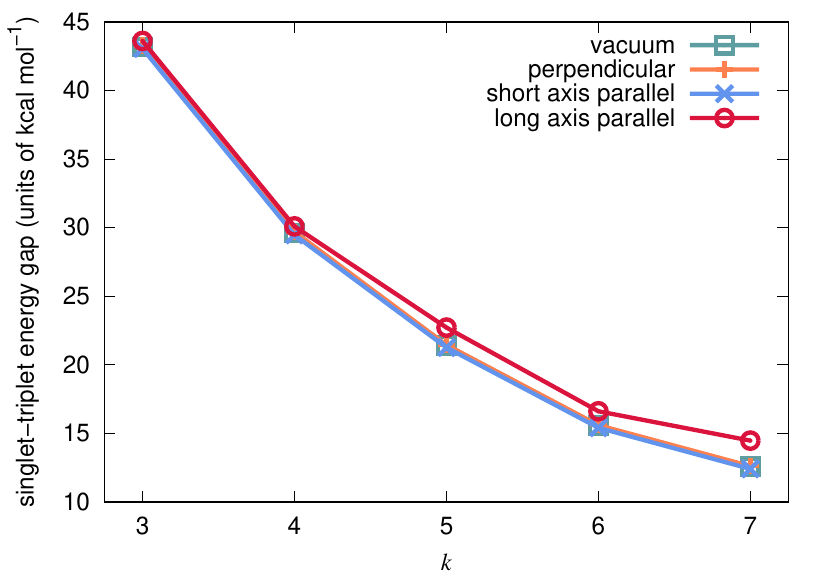}
    \caption{Singlet-triplet energy gaps (units of kcal mol$^{-1}$) for $k$-acene molecules as a function of $k$.} 
    \label{FIG:SINGTRIPOLIGO}
\end{figure}

Figure \ref{FIG:SINGTRIPOLIGO} depicts singlet-triplet energy gaps in $k$-acene molecules as a function of $k$, with cavity parameters $\omega_{\rm cav} = 2.214$ eV and $\lambda = 0.05$ atomic units.   All data presented here were obtained using an active-space-based representation of the electronic structure of the oligoacene molecules; details regarding the active space specification can be found in Sec.~\ref{SEC:COMPUTATION_DETAILS}. We consider isolated molecules (labeled ``vacuum'' in Fig.~\ref{FIG:SINGTRIPOLIGO}) and cavity-coupled molecules in three orientations relative to the cavity mode polarization axis, where (i) the plane of the molecule is perpendicular to the polarization direction (``perpendicular''), (ii) the short axis of the molecule is parallel to the polarization direction (``short axis parallel''), or (iii) the long axis of the molecule is parallel to the polarization direction (``long axis parallel''). First, we note that, for all four cases, the singlet-triplet gap decreases monotonically with increasing $k$, as expected.\cite{DePrince16_2260,Mazziotti08_134108,Chan07_134309} Second, for two orientations of cavity-bound molecules (perpendicular and short axis parallel), the singlet-triplet gaps are difficult to distinguish from those for the isolated species, at least on the scale of Fig.~\ref{FIG:SINGTRIPOLIGO}. On the other hand, when the long axis of the molecule is parallel to the mode polarization axis, the singlet-triplet gap widens noticeably, which indicates that the cavity destabilizes the triplet states, relative to the singlet states.  For 5-, 6-, and 7-acene in this orientation, the singlet-triplet energy gaps increase by 
1.3, 1.1, and 1.9 kcal mol$^{-1}$, 
respectively, relative to the values obtained in the isolated-molecule limit; this increase is roughly 15\% in the case of 7-acene. 


\begin{figure*}
    \centering
    \includegraphics{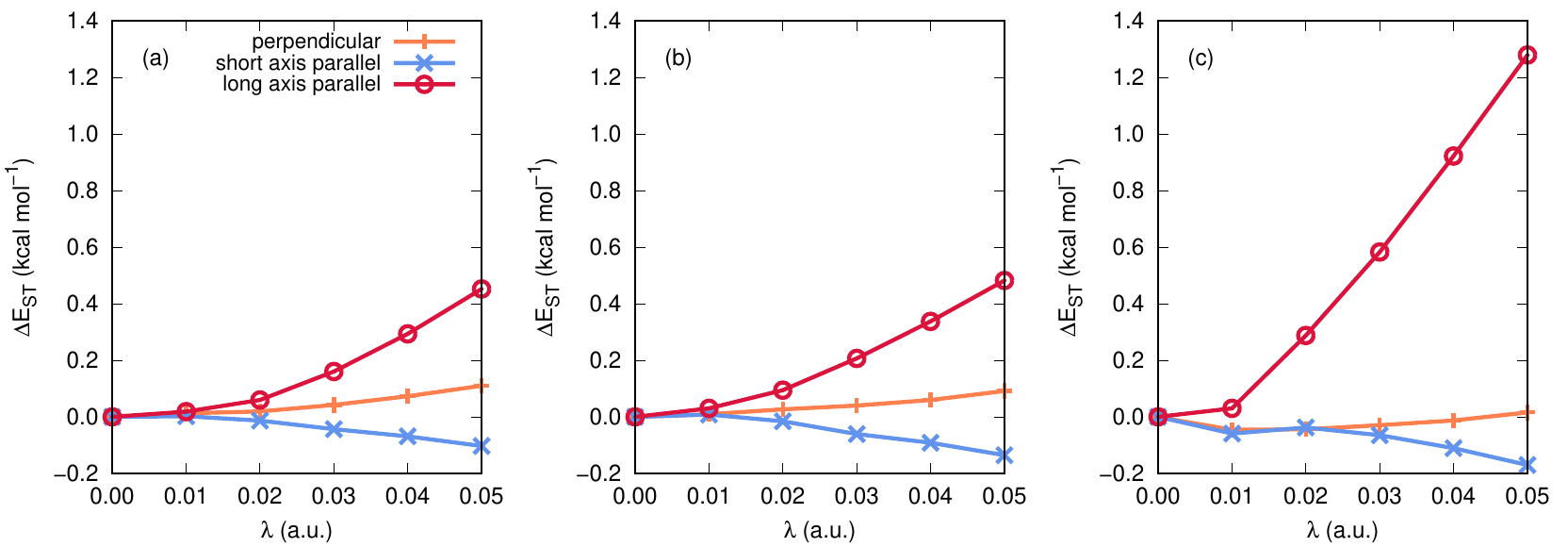}
    \caption{Changes to the singlet-triplet energy gap (units of kcal mol$^{-1}$) for (a) 3-acene, (b) 4-acene, and (c) 5-acene as a function of cavity coupling strength.}
    \label{FIG:SINGTRIPLAMBDA}
\end{figure*}

We explore the orientational dependence of the singlet-triplet energy gaps further in Fig.~\ref{FIG:SINGTRIPLAMBDA}, which depicts cavity-induced changes to these gaps (denoted $\Delta E_{\rm ST}$) for 3-, 4-, and 5-acene as a function of cavity coupling strength. Here, the cavity mode frequency is fixed at 2.214 eV. As in Fig.~\ref{FIG:SINGTRIPOLIGO}, we consider three different molecular orientations, and we observe different behavior in $\Delta E_{\rm ST}$ depending on the orientation. First, the molecules are least sensitive to the presence of the cavity when the plane of the molecule is oriented perpendicular to the cavity mode polarization; the largest observed change in this orientation is a 
0.1 kcal mol$^{-1}$ increase in the gap for 3-acene, at a coupling strength of 0.05 atomic units. The molecules are slightly more sensitive to cavity interactions when the short axis of the molecule is parallel to the mode polarization direction. In this case, we observe a general decrease in the singlet-triplet energy gap with increasing coupling strength; the largest observed $\Delta E_{\rm ST}$ value for this orientation was 
-0.2 kcal mol$^{-1}$ for 5-acene, at a coupling strength of 0.05 atomic units. 
Note that the magnitudes of these changes are not much larger than the convergence criterion used for the primal-dual energy gap in these calculations (10$^{-4}$ E$_{\rm h}$ $\approx 0.06$ kcal mol$^{-1}$), so we can only confidently state that cavity effects are small in these orientations. On the other hand, as was observed in Fig.~\ref{FIG:SINGTRIPOLIGO}, the molecules are much more sensitive to the presence of the cavity when their long axis is oriented along the cavity mode polarization axis. In this orientation, $\Delta E_{\rm ST}$ monotonically increases with increasing coupling strength, for all three molecules. The largest change in the singlet-triplet energy gap we observe here is a 
1.3 kcal mol$^{-1}$ increase for 5-acene, at a coupling strength of 0.05 atomic units. 

For ground states of non-polar cavity-bound molecules ({\em e.g.}, the oligoacene molecules) described by the Hamiltonian given in Eq.~\ref{eq:Hfullcoherent}, the dominant cavity-induced effect stems from the dipole self-energy term. As such, the orientational trends depicted in Fig.~\ref{FIG:SINGTRIPLAMBDA} can be understood in terms of the expectation value of the operator $[\hat{\mu}_{\rm e} - \langle \mu_{\rm e} \rangle]_\xi^2$, for $\xi \in x, y, z$. We have evaluated this expectation value at the v2RDM level of theory, in the absence of any cavity interactions, for 3-, 4-, and 5-acene (see Table \ref{TAB:DIPOLE_VARIANCE}). Here, the short and long axes of the molecules are taken to be along the $y$- and $z$-directions, respectively, and we remind the reader that $\langle \mu_{\rm e} \rangle$ represents the average electric dipole moment evaluated at the Hartree-Fock level of theory. We make the following observations. First, we find that the expectation value of $[\hat{\mu}_{\rm e} - \langle \mu_{\rm e} \rangle]_\xi^2$ has its maximal value for $\xi =z $ and its minimal value for $\xi = x$, for both the singlet and triplet states. This result suggests that the ground states of the oligoacene molecules should be most sensitive to the presence of an optical cavity when their long axis is parallel to the cavity mode axis and least sensitive when they are oriented perpendicular to this axis. Second, the difference in the expectation value of $[\hat{\mu}_{\rm e} - \langle \mu_{\rm e} \rangle]_\xi^2$ for the singlet and triplet states is largest for the $\xi = z$ orientation, which is consistent with the observation that the singlet-triplet gap is most sensitive to the cavity when the long axis is oriented parallel to the cavity mode axis (see Figs.~\ref{FIG:SINGTRIPOLIGO} and \ref{FIG:SINGTRIPLAMBDA}). Third, the sign of the difference in the expectation value of $[\hat{\mu}_{\rm e} - \langle \mu_{\rm e} \rangle]_\xi^2$ for the singlet and triplet states differs for the $\xi = y$ and $\xi = z$ directions, which is consistent with the observation that the singlet-triplet gap opens when the long axis is parallel to the cavity mode axis and closes (slightly) when the short axis is parallel to the cavity mode axis (see Fig.~\ref{FIG:SINGTRIPLAMBDA}).

\begin{table*}[!htpb]
    \caption{The expectation value of the operator $[\hat{\mu}_{\rm e} - \langle \mu_{\rm e} \rangle]_\xi^2$ for $\xi \in x, y, z$, computed at the v2RDM level of theory, in the absence of any cavity interactions, for 3-, 4-, and 5-acene (in atomic units, $e^2 a_0^2$). Values corresponding to the lowest-energy singlet (S) and triplet (T) states are provided, as well as the difference between these values.}
    \label{TAB:DIPOLE_VARIANCE}
    \begin{center}
    \begin{tabular}{lccccccccccc}
        \hline\hline
 & \multicolumn{3}{c}{S} & ~&  \multicolumn{3}{c}{T} &~& \multicolumn{3}{c}{T - S} \\
 \cline{2-4} \cline{6-8} \cline{10-12}
  & x & y & z && x & y & z && x & y & z \\
  \hline
    3-acene  & 47.36 & 53.59 & 57.82 && 47.43 & 53.58 & 58.09 &&  0.07 & -0.01 & 0.27 \\
    4-acene  & 60.28 & 68.00 & 75.93 && 60.38 & 67.94 & 76.92 &&  0.10 & -0.05 & 0.98 \\
    5-acene  & 73.21 & 82.34 & 94.78 && 73.19 & 82.25 & 96.45 && -0.01 & -0.07 & 1.31 \\
    
    \hline
 
\hline\hline
\end{tabular}
\end{center}
\end{table*}

\begin{figure}
    \centering
    \includegraphics{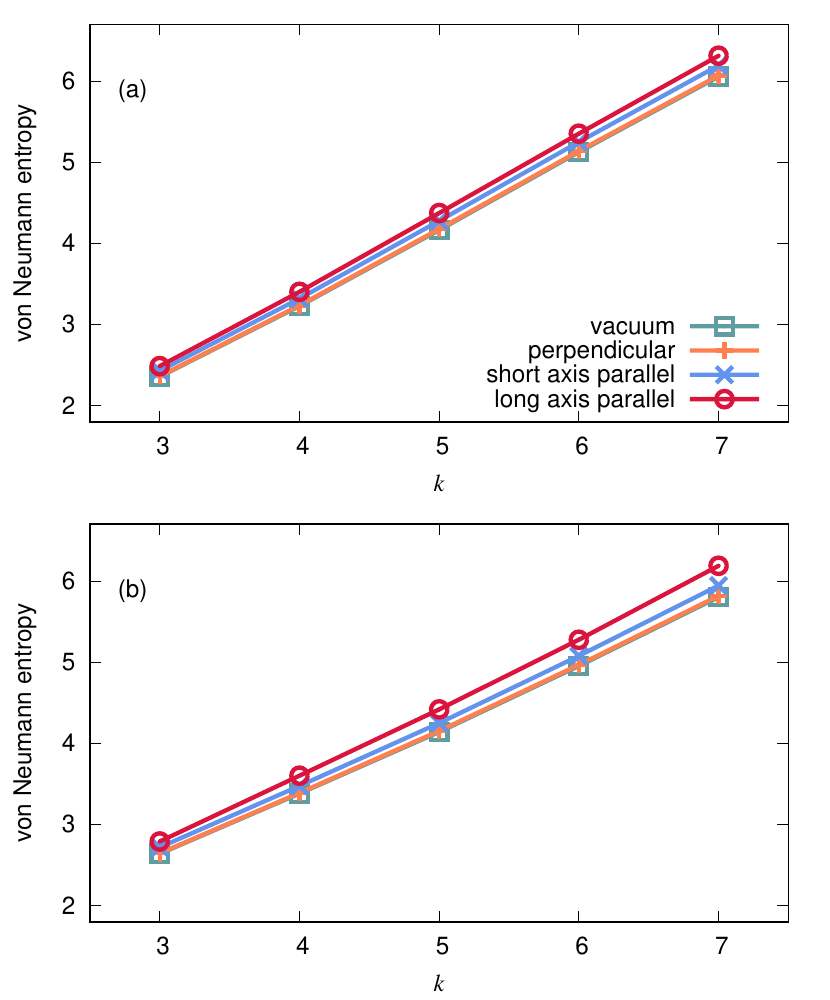}
    \caption{von Neumann entropy associated with the lowest-energy (a) singlet and (b) triplet states of the oligoacene series. }
    \label{FIG:ENTROPY}
\end{figure}

We also evaluated cavity-induced changes to QED-v2RDM-derived natural orbital occupation numbers, which carry information regarding the onset of polyradical character in the longer members of the oligoacene series.  Generally speaking, cavity interactions result in only minor changes to the the natural orbital occupation numbers in both the lowest-energy singlet or triplet states of these molecules. However, subtle changes can be elucidated with the von Neumann entropy,\cite{Ingemar17_Book,VonNeumann18_Book,White06_519,Reiher17_2110} 
\begin{equation}
    S(\eta) = \sum_p \eta_p {\rm ln} \eta_p
\end{equation}
where $\eta_p$ represents a natural orbital occupation number (the eigenvalues of the one-electron RDM). Figure \ref{FIG:ENTROPY} illustrates the von Neumann entropy calculated for isolated oligoacene molecules, as well as three orientations of the cavity-bound species. We find that, for the perpendicular orientation, $S(\eta)$ is essentially unchanged from the isolated-molecule value, for all members of the series. However, noticeable increases in $S(\eta)$ can be seen for the other two orientations, with the larger changes occurring when the long axis is parallel to the cavity mode polarization axis. We observe similar behavior in both the singlet and triplet states, although the changes associated with the triplet state when oriented with the long axis of the molecules parallel to the cavity mode are slightly larger. Because the von Neumann entropy can be interpreted as a measure of electron correlation, these results suggest that strong light-matter coupling enhances electron-electron correlations in oligoacene molecules, and this enhancement is greater in the case of the triplet state. We note that this observed cavity-induced increase in von Neumann entropy is a general phenomenon, not limited to the oligoacene molecules. Indeed, we have also evaluated the von Neumann entropy for the model systems considered in Figs.~\ref{FIG:H2SINGLET} and \ref{FIG:HFSINGLET}, and we find that the cavity-molecule interactions lead to a slight increase of the von Neumann entropy, in general ({\em i.e.}, for all frequencies, bond lengths, and coupling strengths). For example, for H$_2$ (described by the cc-pVDZ basis and at an H--H distance of 0.74 \AA), the von Neumann entropy increases from a cavity-free value of 0.210 to 0.214 when $\lambda = 0.05$. Similarly, for hydrogen fluoride (described by the STO-3G basis and at an H--F distance of 0.917 \AA) respectively, the von Neumann entropy increases from a cavity-free value of 0.153 to 0.156 when $\lambda = 0.05$. Again, for ground states of the Pauli-Fierz Hamiltonian, the dominant cavity effects stem from the dipole self-energy contribution; we can view this term as a repulsive interaction that, in general, leads to slightly stronger correlations among electrons. 

\subsection{Linear hydrogen chains}

Lastly, we apply QED-v2RDM to the symmetric dissociation of linear chains of hydrogen atoms, which is another popular strong-correlation benchmark problem in quantum chemistry.\cite{HeadGordon02_4462,Chan06_144101,Scuseria09_121102,Mazziotti10_014104,Palmieri12_1606,Ayers12_134110,Ma13_224105,Zhang17_031059,Evangelista20_104108,DePrince21_174110} Specifically, we consider the effect that ultra-strong coupling has on the onset of the insulating phase of these systems with increasing H--H separation. As a proxy for this transition, we consider the decay in off-diagonal elements of the one-electron RDM. Previous work has demonstrated that methods designed to model strong electron correlations correctly capture the decay in these elements, while mean-field methods or dynamical correlation models incorrectly predict that the electrons remain delocalized, even at large H--H distances.\cite{Scuseria09_121102,Mazziotti10_014104} Specifically, we consider the atomic-orbital basis representation of the one-electron RDM and the magnitude of the element corresponding to hydrogen atoms at opposite ends of the chains, and we label this quantity $\gamma$. 

\begin{figure}[!htpb]
    \centering
    \includegraphics{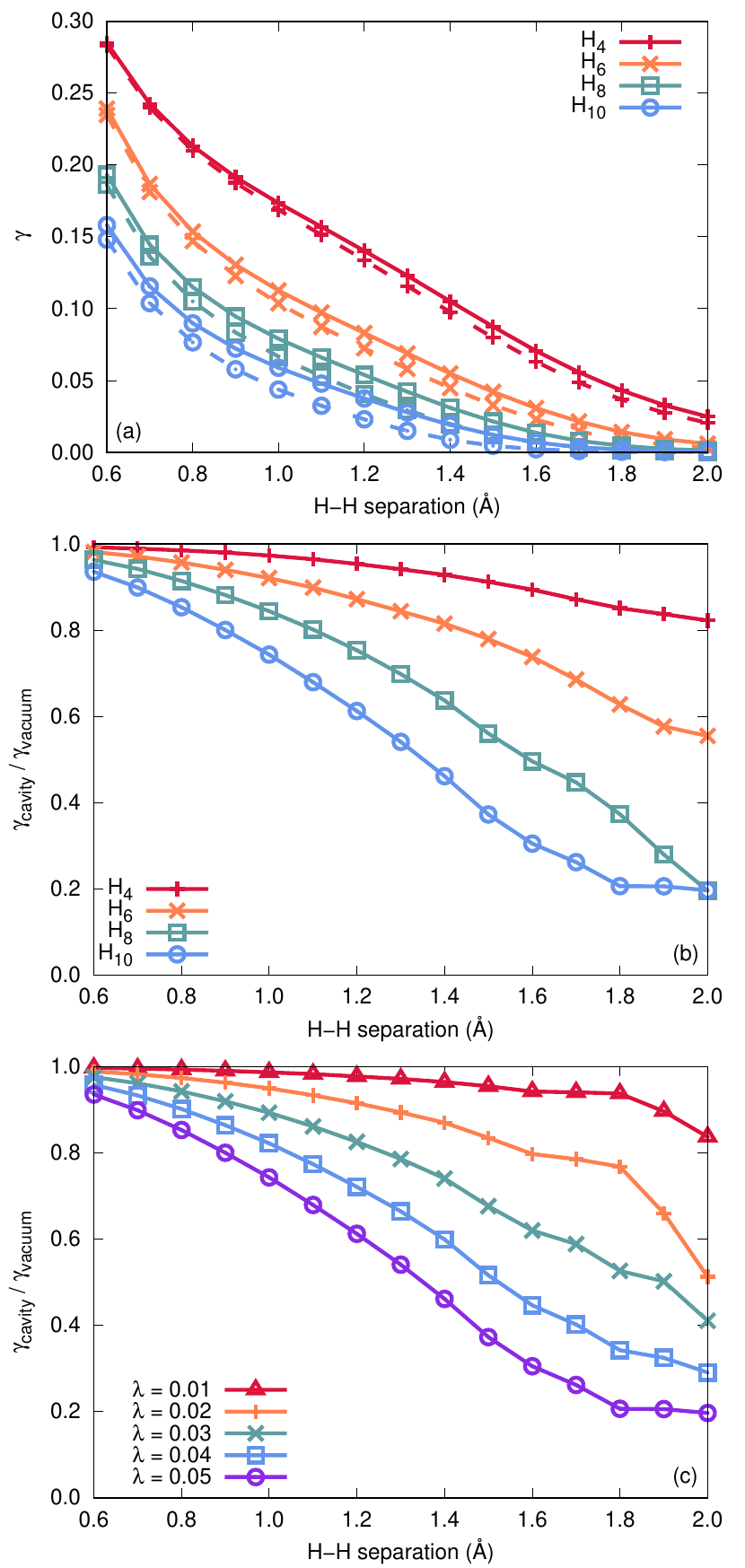}
    \caption{(a) $\gamma$ values for linear hydrogen chains of different lengths, (b) the ratio of $\gamma$ values for isolated and cavity-bound linear hydrogen chains of different lengths, and (c) the ratio of $\gamma$ values for isolated and cavity-bound H$_{10}$ at different cavity coupling strengths. In (a), solid and dashed lines represent isolated and cavity-bound chains, respectively.}
   \label{FIG:HCHAINS}
\end{figure}

Figure \ref{FIG:HCHAINS}(a) depicts $\gamma$ as a function of H--H distance for H$_4$, H$_6$, H$_8$, and H$_{10}$ chains. Here, solid lines represent $\gamma$ values evaluated in the absence of a cavity, while dashed lines refer to cavity-coupled systems with $\omega_{\rm cav}=2.214$ eV, $\lambda=0.05$ atomic units, and the cavity mode polarized along the molecular axis. In all cases, $\gamma$ decreases monotonically with increasing H--H distance, and cavity-coupled chains of a given length consistently exhibit $\gamma$ values that are smaller than those obtained for isolated chains of the same length. Hence, we can conclude that the presence of the cavity increases the insulating character of these systems. Figure \ref{FIG:HCHAINS}(b) provides the ratio of the $\gamma$ values for cavity-coupled and isolated chains, using the same cavity parameters used in panel (a). Here, it becomes clear that cavity-induced changes to the insulating character of the systems become more dramatic for longer chains. 
For example, near the equilibrium geometry (at an H--H distance of 1.0 \AA), $\gamma$ values calculated for cavity coupled systems
fall to 97\%, 92\%, 84\%, and 74\% of the corresponding isolated-chain values for H$_4$, H$_6$, H$_8$, and H$_{10}$, respectively. This trend becomes more apparent at stretched geometries (further into the insulating regime); for example, at an H--H distance of 1.4 \AA, these percentages are 93\%, 82\%, 64\%, and 46\%. Figure \ref{FIG:HCHAINS}(c) illustrates the ratio of the $\gamma$ values for cavity-coupled and isolated chains for H$_{10}$ when varying the cavity coupling strength with fixed mode frequency (again, $\omega_{\rm cav}$ = 2.214 eV). We find that strong light-matter coupling increases the insulating character of H$_{10}$, for all coupling strengths considered and that, not surprisingly, this effect is greatest for the largest coupling strength.


\section{Conclusions}

\label{SEC:CONCLUSIONS}

We have developed a formulation of v2RDM theory that incorporates quantized electromagnetic fields and offers a first-principles description of strong light-matter interactions that complements other recently developed {\em ab initio} QED methods. Unlike QEDFT and QED-CC theories, QED-v2RDM is specifically designed for the simultaneous description of strong electron-electron and strong electron-photon interactions. The formalism also offers computational advantages over other approaches for strong electron correlations. For example, QED-v2RDM can be applied to much larger active spaces than can be treated with a QED generalization of full configuration interaction; we demonstrated this capability by applying the approach to cavity-coupled systems with active spaces as large as (30e,30o).

We derived a set of necessary $N$-representability conditions for polaritonic RDMs that are generalizations of the two-body (PQG) conditions enforced in conventional (non-QED) applications of v2RDM theory. Illustrative calculations on cavity-coupled two-electron and two-hole systems confirmed the exactness of QED-v2RDM for these problems and also revealed the important role of the three-body $N$-representability conditions derived in this work. Additional applications on two classic strong correlation problems suggest that ultra-strong light-matter coupling can be leveraged to realize non-negligible changes to ground-state molecular properties. First, we demonstrated that singlet-triplet energy gaps in oligoacene molecules could be altered by as much as 15\% (in the case of a 7-acene molecule at experimentally realizable\cite{Baumberg18_7146} coupling strengths). Second, we provided clear computational evidence that strong light-matter coupling increases the insulating character of linear chains of hydrogen atoms. These two observations have potential implications for technologies that harness strong light-matter interactions at or below the nanoscale.

\vspace{0.5cm}

\begin{acknowledgments}
This material is based upon work supported by the National Science Foundation under Grants No.~CHE-1554354 and CHE-2100984.\\ 
\end{acknowledgments}

\section{Appendix A: Spin-dependent polaritonic representability conditions}

As mentioned in Sec.~\ref{SEC:THEORY}, various spin symmetries lead to spin-specific polaritonic representability conditions that apply to the RDMs considered in this work. For example, consider a wave function that is an eigenfunction of number operators corresponding to the number of $\alpha$- and $\beta$-spin electrons, {\em i.e.},
\begin{equation}
\label{EQN:NA}
    \hat{N}_\sigma|\Psi\rangle = \sum_{r} \hat{a}^\dagger_{r_\sigma} \hat{a}_{r_\sigma} |\Psi \rangle = N_\sigma| \Psi \rangle
\end{equation}
for $\sigma \in \{\alpha, \beta\}$. Here, we have introduced spin labels for the electronic orbitals, and we assume that the orbitals $p_\alpha$ and $p_\beta$ share a common spatial component. A wave function that satisfies Eq.~\ref{EQN:NA} should also be an eigenfunction of $\hat{S}_z = \frac{1}{2}(\hat{N}_\alpha - \hat{N}_\beta)$, and thus, the magnetic spin quantum number, $M_S = \frac{1}{2} (N_\alpha - N_\beta)$, is expected to be conserved by the RDMs. In addition to the representability conditions outlined in Sec.~\ref{SEC:THEORY}, these RDMs should satisfy contraction constraints that arise by projecting Eq.~\ref{EQN:NA} onto bra states defined by $\langle \Psi | \hat{a}_{p_\tau}^\dagger \hat{a}_{q_\tau}$, leading to
\begin{equation}
    \sum_{r} {}^2D^{p_\tau r_\sigma}_{q_\tau r_\sigma} = (N_\sigma - \delta_{\sigma \tau}) {}^1D^{p_\tau}_{q_\tau}
\end{equation}
which is a spin-dependent analogue of Eq.~\ref{eq:12relate}. Similarly, projecting Eq.~\ref{EQN:NA} onto bra states defined by $\langle \Psi|\hat{D}_{AB}$ and $\langle \Psi | \hat{a}_{p_\tau}^\dagger \hat{a}_{q_\tau} \hat{D}_{AB}$ leads to spin-dependent analogues of Eqs.~\ref{eq:12relatephoton} and \ref{eq:123interrelate}, respectively. It can also be shown that RDMs deriving from a wave function that satisfies both Eq.~\ref{EQN:NA} and Eq.~\ref{EQN:S2} should satisfy
\begin{equation}
\sum_{pq} {}^3D^{p_\alpha q_\beta A}_{q_\alpha p_\beta B} = \bigg [ \frac{1}{2}N + M_S^2 - S(S+1) \bigg ] {}^1D^A_B
\end{equation}
Lastly, we consider RDMs that correspond to maximal spin-projection states ($M_S = S$) that satisfy
\begin{equation}
\label{EQN:S+}
    \hat{S}_+ |\Psi \rangle = \sum_r \hat{a}^\dagger_{r_\alpha} \hat{a}_{r_\beta}|\Psi\rangle = 0
\end{equation}
This equation implies a set of constraints on the particle-hole RDM\cite{Ayers12_014110} that can be generalized to the polaritonic case by projecting Eq.~\ref{EQN:S+} onto bra states defined by $\langle \Psi |\hat{a}^\dagger_{p_\beta}\hat{a}_{q_\alpha} \hat{D}_{AB}$ to give 
\begin{equation}
    \sum_r {}^3G^{p_\beta q_\alpha A}_{r_\beta r_\alpha B} = 0
\end{equation}

\section{Appendix B: Additional data}

\subsection{Potential energy curves for molecular hydrogen}

\begin{figure*}[!hpb]
    \centering
    \includegraphics{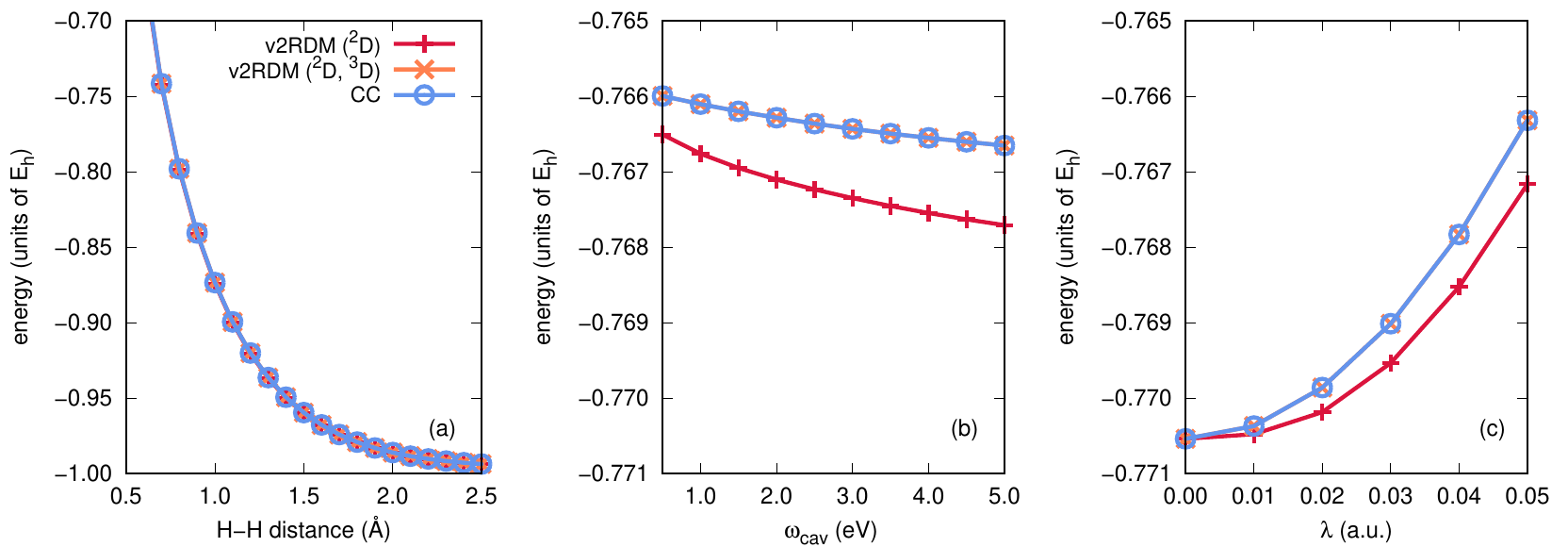}
    \caption{Potential energy curves (units of E$_{\rm h}$) as a function of (a) interatomic distance, (b) cavity mode frequency, and (c) cavity coupling strength for the lowest-energy triplet state of molecular hydrogen described by the cc-pVDZ basis set.}
    \label{FIG:H2TRIP}
\end{figure*}

Figure \ref{FIG:H2TRIP} illustrates potential energy curves for the lowest-energy triplet state of a two-electron system, H$_2$, evaluated at the QED-CCSD-1 and QED-v2RDM levels of theory, using the cc-pVDZ basis set. The molecule is coupled to a single-mode cavity with the cavity mode polarization axis parallel to the molecular axis. Panel (a) of Fig.~\ref{FIG:H2TRIP} depicts the energy as a function of the H--H distance, with cavity parameters $\omega_{\rm cav}$ and $\lambda$ fixed at 2.214 eV and 0.05, respectively. Panel (b) depicts a scan over $\omega_{\rm cav}$, with fixed H--H distance (0.74 \AA) and fixed coupling strength ($\lambda = 0.05$). Lastly, panel (c) provides energies computed at various coupling strengths, with fixed H--H distance (0.74 \AA) and fixed cavity mode frequency (2.214 eV). 
Two curves in each panel are labeled v2RDM, and these curves differ in the $N$-representability conditions applied within the computations. The label ``v2RDM (${}^2$D)'' indicates that the calculations were carried out under conditions that depend upon up to two-body RDMs (specifically, the one-electron, one-hole, two-electron, photon,  and electron-photon RDMs), while the label ``v2RDM (${}^2$D, ${}^3$D)'' indicates the application of the same conditions plus those that depend on the electron-electron-photon RDM. The definitions of these RDMs can be found in Table I of the main text. 

\subsection{Absolute energies for linear oligoacene molecules}

Table \ref{TAB:POTENERGY} provides QED-v2RDM derived energies for the lowest-energy singlet and triplet states of the oligoacene series ($k$-acene, $k$=3--5). Results are provided for molecules in three orientations relative to the cavity mode polarization axis at various coupling strengths. In all cases $\omega_{\rm cav}$ = 2.214 eV. Differences between the singlet and triplet energies correspond to the singlet-triplet energy gaps illustrated in Figs.~3 and 4 of the main text. Active space details can be found in the main text.

\begin{table*}[!hpb]
    \caption{QED-v2RDM-derived energies (E$_h$) of the lowest-energy singlet and triplet $k$-acene molecules $k=3-5$ for different cavity coupling strengths $\lambda$ in three different orientations with respect to the cavity mode polarization axis.}
    \label{TAB:POTENERGY}
    \begin{center}
    \begin{tabular}{l c c c c c c c c}
    \hline\hline
    & & \multicolumn{3}{c}{singlet} &~& \multicolumn{3}{c}{triplet} \\                                        
    \cline{3-5} \cline{7-9}
    orientation & $\lambda$ & 3-acene & 4-acene & 5-acene &~& 3-acene & 4-acene & 5-acene\\
    \hline
    \hline
    vacuum        & -    & -536.2106 & -688.9204 & -841.6307 &~& -536.1418 & -688.8732 & -841.5965 \\
        \hline
    perpendicular & 0.01 & -536.2083 & -688.9175 & -841.6270 &~& -536.1395 & -688.8702 & -841.5929 \\
                  & 0.02 & -536.2014 & -688.9086 & -841.6163 &~& -536.1325 & -688.8614 & -841.5822 \\ 
                  & 0.03 & -536.1898 & -688.8940 & -841.5985 &~& -536.1210 & -688.8467 & -841.5644 \\
                  & 0.04 & -536.1737 & -688.8734 & -841.5736 &~& -536.1048 & -688.8261 & -841.5394 \\
                  & 0.05 & -536.1529 & -688.8470 & -841.5416 &~& -536.0840 & -688.7997 & -841.5074 \\
    \hline
    short axis parallel~~~~ & 0.01  & -536.2080 & -688.9171 & -841.6266 &~& -536.1392 & -688.8698 & -841.5925 \\
                            & 0.02  & -536.2001 & -688.9071 & -841.6146 &~& -536.1314 & -688.8600 & -841.5805 \\ 
                            & 0.03  & -536.1872 & -688.8907 & -841.5948 &~& -536.1184 & -688.8436 & -841.5607 \\
                            & 0.04  & -536.1692 & -688.8680 & -841.5674 &~& -536.1005 & -688.8210 & -841.5334 \\
                            & 0.05  & -536.1464 & -688.8392 & -841.5326 &~& -536.0777 & -688.7922 & -841.4987 \\
    \hline
    long axis parallel & 0.01 & -536.2078  & -688.9167 & -841.6260 &~& -536.1390 & -688.8694 & -841.5917 \\
                       & 0.02 & -536.1994  & -688.9058 & -841.6126 &~& -536.1305 & -688.8584 & -841.5780 \\ 
                       & 0.03 & -536.1857  & -688.8882 & -841.5911 &~& -536.1167 & -688.8407 & -841.5560 \\
                       & 0.04 & -536.1670  & -688.8642 & -841.5619 &~& -536.0977 & -688.8165 & -841.5263 \\
                       & 0.05 & -536.1433  & -688.8339 & -841.5249 &~& -536.0738 & -688.7859 & -841.4887 \\
\hline\hline
\end{tabular}
\end{center}
\end{table*}

\bibliography{Journal_Short_Name.bib,rdm.bib,cqed.bib,other.bib}

\end{document}